\documentclass[aps,showpacs,preprintnumbers,amsmath,amssymb]{revtex4}
\UseRawInputEncoding
\usepackage{dcolumn}
\usepackage{multirow}
\usepackage[dvips]{epsfig}
\def \be {\begin{equation}}
\def \ee {\end{equation}}
\def \bea {\begin{eqnarray}}
\def \eea {\end{eqnarray}}

\begin{document}
\baselineskip=0.8 cm
\title{\bf Effect of noncircularity on the dynamic behaviors of particles in a disformal rotating black-hole spacetime }
\author{Xuan Zhou $^{1}$, Songbai Chen$^{1,2}$\footnote{Corresponding author: csb3752@hunnu.edu.cn}, Jiliang Jing$^{1,2}$ }
%\email{csb3752@163.com}

\affiliation{ $ ^1$ Department of Physics,  Synergetic Innovation Center for Quantum Effects and Applications, Hunan
Normal University,  Changsha, Hunan 410081, People's Republic of China
\\
$ ^2$Center for Gravitation and Cosmology, College of Physical Science and Technology, Yangzhou University, Yangzhou 225009, People's Republic of China}

\begin{abstract}
\baselineskip=0.6 cm
\begin{center}
{\bf Abstract}
\end{center}

A disformal rotating black-hole solution is a black-hole solution in quadratic degenerate higher-order scalar-tensor theories. It breaks the circular condition of spacetime different from the case of the usual Kerr spacetime. This study investigated the dynamic behaviors of the motion of timelike particles in such disformal black-hole spacetime with an extra deformation parameter. Results showed that the characteristics of the particle's motion depend on the sign of the deformation parameter. For the positive deformation parameter, the motion is regular and orderly. For the negative one, as the deformation parameter changes, the motion of the particles undergoes a series of transitions between the chaotic motion and the regular motion and falls into the horizon or escapes to spatial infinity. This means that the dynamic behavior of timelike particles in the disformal Kerr black-hole spacetime with noncircularity becomes richer than that in the usual Kerr black-hole case.

\end{abstract}


\maketitle
\newpage

\section{Introduction}

Geodesics around black holes have been extensively studied as it could help understand the properties of black holes, the geometric structure of spacetimes, and the motion of particles. Especially, timelike geodesics with a back reaction can also be applied to simulate the inspiral of black-hole binaries with extreme mass ratio due to gravitational wave emission \cite{emris1,emris2,emris3,emris4,emris5}. Negative energy geodesics are always used to estimate the energy absorbed by particles passing through the ergoregion of a rotating black hole using the Penrose process \cite{Penrose1,Penrose2}. Moreover, null geodesics are also relevant for studying the shadow of a black hole that was observed in the direct imaging of the supermassive black hole M87* by the Event Horizon Telescope \cite{bhshaw}.
In general relativity, it is expected that an astrophysical black hole in the universe should be a Kerr black hole in terms of the well-known no-hair theorem. The geodesic motion of particles in the Kerr black-hole spacetime is integrable because the geodesic equation is variable-separable, and the number of integrals of motion in the dynamic system is equal to its degree of freedom \cite{Carter}. Then, the geodesic motion of particles is regular and orderly in such black-hole spacetimes. However, in some alternative theories of gravity, the geometries of black holes could become more complicated so that the geodesic equation is not variable-separable, and chaos may emerge in the geodesic motion of particles.
One of the fascinating alternative theories of gravity is the so-called degenerate higher-order scalar-tensor (DHOST) theories \cite{Langlois,Langlois1, Crisostomi,Achour1,Achour2}, which contain the higher-order derivative of the scalar field and satisfy with a certain set of degeneracy conditions.
For a higher derivative theory, some auxiliary variables are introduced to replace the higher-order time derivatives of the Lagrangian with first-order time derivatives. Then, a so-called kinetic matrix \cite{Langlois, Achour1} is obtained using the coefficients of the terms quadratic in time derivatives. If the kinetic matrix is invertible, the system includes the extra degree of freedom associated with the Ostrogradsky ghost. The so-called degeneracy conditions are conditions that make the kinetic matrix degenerate (i.e., the matrix determinant is zero) so that the extra degree of freedom related to ghosts can be eliminated. In DHOST theories, the degeneracy conditions on Lagrangian ensure that the Ostrogradsky ghost is absent even if there exist higher-order equations of motion, which means that the degeneracy of Lagrangian is crucial for higher-order theories with only a single scalar degree of freedom \cite{Langlois,Langlois1}. In general alternative theories of gravity, it is difficult to obtain exact solutions for black holes because of the more complicated field equations. However, in DHOST theories, some new black-hole solutions were obtained recently \cite{Crisostomi,Achour1,Achour2,Babichev,Charmousis,Aelst}. These solutions can be classified into two types. One is the so-called stealth solution, whose metric is the same as those in general relativity, and the extra scalar field does not emerge in the spacetime metric. The other is the nonstealth solution, where the parameters of the scalar field appear in the metric, and its metric form deviates from that in Einstein's theory.
DHOST theories can be classified in terms of the concrete forms of degeneracy conditions \cite{Langlois,Achour3}, which depend on the functions $A_i$ $(i=1,2...5)$ and $F$ in the action (\ref{act1}). The quadratic DHOST Ia theory \cite{Langlois,Achour3} is characterized by the property $A_1+A_2=0$ and $F+XA_2\neq0$ (where $X$ is the kinetic term of the scalar field in the DHOST theory). The usual quartic Horndeski theory is a special case in the quadratic DHOST Ia theory \cite{Langlois,Achour3}.
Interestingly, disformal and conformal transformations can lead from some DHOST Ia theory to some other specific DHOST Ia theory \cite{Achour1}. Thus, when a disformal transformation is made on a known "seed" metric $\tilde{g}_{\mu \nu}$ in the DHOST Ia theory, a new solution $g_{\mu \nu}$ can be obtained in another specific DHOST Ia theory. With such a technique, a disformal rotating black-hole solution in quadratic DHOST theories was obtained recently \cite{Achour3,Anson}.
This disformal solution is nonstealth and has three parameters. The mass, the spin, and the deformation parameter described the deviation from the Kerr geometry. The scalar field attached to the disformal solution is time-dependent with a constant kinetic density. Although it is a nonstealth solution, the disformed Kerr solution is still asymptotically flat and has a single curvature singularity as in the usual Kerr case. However, the scalar field makes the disformal spacetime no longer Ricci flat. Especially, the presence of the metric function $g_{rt}$ also leads to the noncircularity \cite{circle1,circle2,circle3} in the disformal Kerr spacetime \cite{Achour3,Anson}. These spacetime properties could modify the geodesic motion of particles and yield new observational effects different from those in the Kerr case. A recent study indicated \cite{LongF} that the shadow of a disformal rotating black hole heavily depends on the deformation parameter, and some eyebrow-like shadows with self-similar fractal structures appear as the deformation parameter lies in certain special ranges. Moreover, the deformation parameter in the disformal Kerr black hole was constrained by quasi-periodic oscillations with the observation data of GRO J1655-40 \cite{Chenwz} and the noncircularity of the spacetime was examined by Sagittarius A* with orbiting pulsars \cite{Takamori}. The post-Newtonian motion of stars orbiting the disformal Kerr black hole was analyzed using the osculating orbit method \cite{Anson2}. This paper aimed to investigate the chaotic motion of particles around the disformal Kerr black hole in DHOST theories.
The paper is organized as follows: Section II briefly introduces a disformal Kerr black-hole solution with noncircularity in quadratic DHOST theories and presents the geodesic equation of a test timelike particle. In Section III, the noncircularity makes the equation of motion not variable-separable, leading to the chaotic phenomenon that occurred in the corresponding particle's dynamic system. With techniques such as the Poincar\'e section, Lyapunov exponents, and bifurcation diagram, this study also investigated the effects of the deformation parameter and the black-hole spin parameter on the chaotic motion for a chosen timelike particle. Finally, this paper ends with a summary.

\section{Geodesic equation of particles around a disformal Kerr black hole in quadratic DHOST theories}
This section briefly introduces a disformal Kerr black-hole solution in quadratic DHOST theories. It belongs to nonstealth rotating solutions due to an extra deformation parameter. The most general action in quadratic DHOST theories can be expressed as \cite{Achour3,Anson}
\begin{equation}
S=\int d^{4} x \sqrt{-g}\left(P(X, \phi)+Q(X, \phi) \square \phi+F(X, \phi) R+\sum_{i=1}^{5} A_{i}(X, \phi) L_{i}\right).\label{act1}
\end{equation}
Here $R$ is the usual Ricci scalar, and the functions $A_{i}$, $F$, $Q$, and $P$ depend on the scalar field $\phi$ and its kinetic term $X \equiv \phi_{\mu} \phi^{\mu}$, where $\phi_{\mu} \equiv \nabla_{\mu} \phi$. $L_{i}$ are the Lagrangians containing quadratic in the second derivatives of the scalar field $\phi$, which are defined by
\begin{equation}\begin{array}{l}
L_{1} \equiv \phi_{\mu \nu} \phi^{\mu \nu}, \quad L_{2} \equiv(\square \phi)^{2}, \quad L_{3} \equiv \phi^{\mu} \phi_{\mu \nu} \phi^{\nu} \square \phi \\
L_{4} \equiv \phi^{\mu} \phi_{\mu \nu} \phi^{\nu \rho} \phi_{\rho}, \quad L_{5} \equiv\left(\phi^{\mu} \phi_{\mu \nu} \phi^{\nu}\right)^{2},
\end{array}\end{equation}
where $\phi_{\mu \nu} \equiv \nabla_{\nu} \nabla_{\mu} \phi$ denotes the second covariant derivatives of $\phi$.
In quadratic DHOST theories (\ref{act1}), to avoid Ostrogradsky instabilities and ensure only an extra scalar degree of freedom besides the usual tensor modes of gravity, the functions $F$ and $A_i$ must satisfy the degeneracy conditions, which result in the kinetic matrix determinant being zero. The zero-value determinant of the kinetic matrix yields an expression of the form \cite{Langlois}
\begin{equation}
D_0(X)+D_1(X)U^2_{*}+D_2(X)U^4_{*}=0,
\end{equation}
where quantity $U_{*}$ is related to the auxiliary field $U_{\mu}=\nabla_{\mu}\phi$. The functions $D_0(X)$, $D_1(X)$, and $D_2(X)$ depend on the six arbitrary functions $F$ and $A_i$. For simplicity, the formulas for $D_0(X)$, $D_1(X)$, and $D_2(X)$ are not listed, and the readers are referred to the literature \cite{Langlois} for the detailed $D_i(X)$. By solving these three conditions $D_i(X)=0$, $i=0,1,2$, all DHOST theories can be classified. The quadratic DHOST I theory \cite{Langlois,Achour3} is characterized by the property $A_1+A_2=0$. The subclass Ia satisfies $F+XA_2\neq0$, which means that the degenerate theories in class Ia depend on three arbitrary functions $A_2$, $A_3$, and $F$. Another subclass (Ib) meets $F+XA_2=0$ and $A_3=2(F-2XF_{X})/X^2$ (where $F_{X}$ is the partial derivative of $F$ with respect to $X$), and the functions $A_4$, $A_5$, and $F$ are arbitrary. The famous quartic Horndeski theory is a special case in the quadratic DHOST Ia theory \cite{Langlois,Achour3}.
It is well known that a new solution in the DHOST Ia theory can be obtained from a "seed" known solution by performing a disformal transformation. The disformal transformation of the metric can be expressed as \cite{Langlois,Achour3}
\begin{equation}\label{conformal}
g_{\mu \nu}=A(\phi, X) \tilde{g}_{\mu \nu}-B(\phi, X) \phi_{\mu} \phi_{\nu},
\end{equation}
where $g_{\mu \nu}$ is the "disformed" metric and $\tilde{g}_{\mu \nu}$ is the original "seed" one. $A(X,\phi)$ and $B(X,\phi)$ are conformal and disformal factors, respectively. To obtain a nonstealth solution, the functions $A$ and $B$ must satisfy certain conditions so that the two metrics are not degenerate. Starting from the usual Kerr metric in general relativity and adopting the transformation with $A(X, \phi)=1$ and $B(X, \phi)=B_{0}$, the disformal Kerr metric in Boyer--Lindquist coordinates \cite{Achour3,Anson} can be obtained.
\begin{equation}\label{metric}
\begin{aligned}
d s^{2}=&-\frac{\Delta}{\rho^2}\left(d t-a \sin ^{2} \theta d \varphi\right)^{2}+\frac{\rho^2}{\Delta} d r^{2}+\rho^2 d \theta^{2}+\frac{\sin ^{2} \theta}{\rho^2}\left(a d t-\left(r^{2}+a^{2}\right) d \varphi\right)^{2} \\
&+\alpha\left(d t + \sqrt{2 M r\left(r^{2}+a^{2}\right)} / \Delta d r\right)^{2},
\end{aligned}
\end{equation}
with
\begin{equation}
\alpha \equiv-B_{0} m^{2},\quad\quad\quad \Delta \equiv r^{2}-2 M r+a^{2}, \quad\quad\quad \rho^{2} \equiv r^{2}+a^{2} \cos ^{2} \theta,
\end{equation}
where $M$ and $a$ are the general black-hole mass and the spin parameter, respectively. $\alpha$ is the deformation parameter of black hole that is related to the rest mass $m$ of the scalar field. The determinant of the metric (\ref{metric}) is $g=(\alpha-1)\rho^4\sin^2\theta$, and the metric becomes degenerate when $\alpha=1$. Moreover, to maintain its Lorentz signature, one must have $\alpha<1$. The choice of $A(X, \phi)=1$ can simplify the metric coefficients in the disformal black-hole solution \cite{Achour3}. Here, the scalar field is taken as only a function of the coordinates $t$ and $r$ with a form \cite{Achour3,Anson}.
\begin{eqnarray}
&&\phi(t, r)=-m t+S_{r}(r), \quad\quad\quad S_{r}= -\int \frac{\sqrt{\mathcal{R}}}{\Delta} d r, \nonumber\\
&&\mathcal{R}=2 M m^{2} r(r^{2}+a^{2}),\quad\quad\quad \Delta=r^{2}+a^{2}-2 M r.\label{scalar}
\end{eqnarray}
Obviously, the scalar field is divergent as $\sqrt{r}$ at large $r$, which is nonetheless a less appealing feature of this solution. However, this is not a physical problem, as the scalar field $\phi$ interacts with gravity only through its gradient \cite{Charmousis} from Eq.(\ref{conformal}) with $A(X, \phi)=1$ and $B(X, \phi)=B_{0}$, and the form of the scalar field (\ref{scalar}) can avoid the pathological behavior of the disformal metric at spatial infinite. Like in the Kerr black-hole case, the disformal Kerr metric (\ref{metric}) also has an intrinsic ring singularity at $\rho=0$, and the spacetime is asymptotically flat. However, note that the disformal Kerr metric (\ref{metric}) is not Ricci flat, i.e., $R_{\mu\nu}\neq0$, which is different from the general Kerr case. Moreover, the $drdt$ term yields the lack of circularity in the disformal Kerr spacetime \cite{Achour3,Anson}, and the spacetime cannot be foliated by two-dimensional meridional surfaces everywhere orthogonal to the Killing field $\xi=\partial_t$ and $\eta=\partial_{\phi}$ \cite{circle1,circle2,circle3}. This is qualitatively different from the usual Kerr spacetime in general relativity. The absence of circularity modifies the structure of black-hole horizons so that the horizons depend on the polar angle $\theta$ and cannot be given by $r=const$ in Boyer--Lindquist coordinates. The corresponding surface gravity is no longer a constant \cite{Achour3,Anson}.
In a curved spacetime, the Lagrangian of a timelike particle moving along the geodesic is
\begin{equation}
\begin{aligned}
\mathcal{L}=\frac{1}{2} g_{\mu\nu}\dot{x}^{\mu}\dot{x}^{\nu}=\frac{1}{2} \left(g_{t t} \dot{t}^{2}+g_{r r} \dot{r}^{2}+g_{\theta \theta} \dot{\theta}^{2}+g_{\varphi \varphi} \dot{\varphi}^{2}+2 g_{t \varphi} \dot{t} \dot{\varphi}+2 g_{tr} \dot{t}\dot{r} \right),\label{lagran}
\end{aligned}
\end{equation}
where the dots denote derivatives with respect to the proper time $\tau$. Obviously, the metric functions in the disformal Kerr spacetime (\ref{metric}) are independent of the coordinates $t$ and $\varphi$. It means that $t$ and $\varphi$ are cyclic coordinates for the Lagrangian (\ref{lagran}), and there exist two conserved quantities for a timelike particle in the disformal Kerr spacetime (\ref{metric}), i.e., the energy $E$ and the $z$-component of the angular momentum $L$,
\begin{eqnarray}
E=-p_{t}=-g_{tt}\dot{t}-g_{tr}\dot{r}-g_{t\varphi}\dot{\varphi}, \quad \quad \quad L=p_{\varphi}=g_{t\varphi}\dot{t}+g_{\varphi\varphi}\dot{\varphi}.\label{conserved quantities}
\end{eqnarray}
With these two conserved quantities, the geodesic equation for the disformal Kerr black-hole spacetime can be written as
\begin{equation}
\dot{t}=\frac{g_{\varphi \varphi} E+g_{t \varphi} L+g_{t r} g_{\varphi\varphi} \dot{r}}{g_{t \varphi}^{2}-g_{t t} g_{\varphi \varphi}},\quad\quad\quad
\dot{\varphi}=\frac{g_{t \varphi} E+g_{t t} L+g_{t r} g_{t \varphi} \dot{r}}{g_{t t} g_{\varphi \varphi}-g_{t \varphi}^{2}}.\label{motion1}
\end{equation}
and
\begin{equation}
\tilde{g}_{r r} \dot{r}^{2}+g_{\theta \theta} \dot{\theta}^{2}=V_{e f f}\left(r, \theta ; E, L\right), \quad \quad \quad \tilde{g}_{r r}=\left[g_{r r}+\frac{g_{t r}^{2} g_{\varphi\varphi}}{g_{t \varphi}^{2}-g_{t t} g_{\varphi \varphi}}\right],\label{motion3}
\end{equation}
with the effective potential
\begin{equation}
V_{e f f}\left(r, \theta ; E, L\right)=\frac{E^{2} g_{\varphi\varphi}+2 E L g_{t \varphi}+L^{2} g_{t t}}{g_{t \varphi}^{2}-g_{t t} g_{\varphi\varphi}}-1.\label{motion4}
\end{equation}
Moreover, the timelike particles in the disformal Kerr spacetime (\ref{metric}) must obey the constraint condition
\begin{equation}
h=g_{t t} \dot{t}^{2}+g_{r r} \dot{r}^{2}+g_{\theta \theta} \dot{\theta}^{2}+g_{\varphi \varphi} \dot{\varphi}^{2}+2 g_{t \varphi} \dot{t} \dot{\varphi}+2 g_{t r} \dot{t} \dot{r}+1=0.\label{con4}
\end{equation}
In the case with the nonzero deformation parameter $\alpha \neq 0$, the dynamic system is nonintegrable because the condition (\ref{con4}) is not variable-separable, and the system admits only two integrals of motion $E$ and $L$. This implies that the motion of the particle could be chaotic in the disformal Kerr black-hole spacetime (\ref{metric}). The next section will investigate the effect of the deformation parameter $\alpha$ on the geodesic motion of a timelike particle around a disformal Kerr black hole (\ref{metric}).

\section{Chaotic motion of timelike particles moving in the disformal rotating Kerr black-hole spacetime}
To probe chaotic behaviors of timelike particles in the disformal Kerr black-hole spacetime (\ref{metric}), numerically differential equations must be solved (\ref{motion1})-(\ref{motion3}). This study adopted the corrected fifth-order Runge--Kutta method wherein high precision can be effectively ensured \cite{Huang,DZMa1,DZMa2,DZMa3}. In the disformal Kerr black-hole spacetime (\ref{metric}), the motion of the particle is entirely determined by the black-hole background parameters $\{M, a, \alpha \}$, the particle's parameters $\{E,L\}$, and the corresponding initial conditions of the particle $\{r,\theta, \dot{r},\dot{\theta}\}$.
\begin{figure}[ht]
\includegraphics[width=5.0cm ]{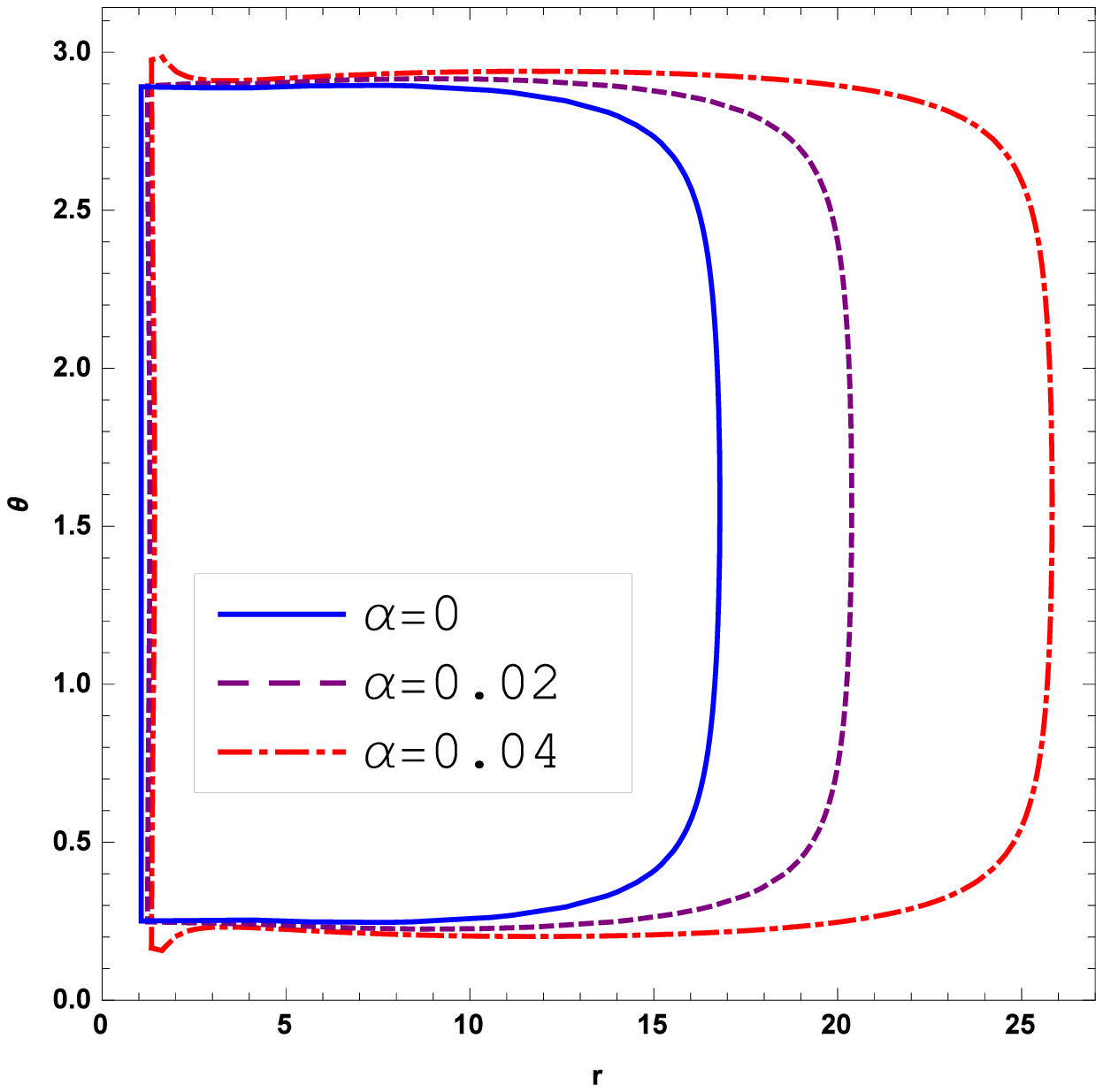}\;\;\;\includegraphics[width=5.0cm ]{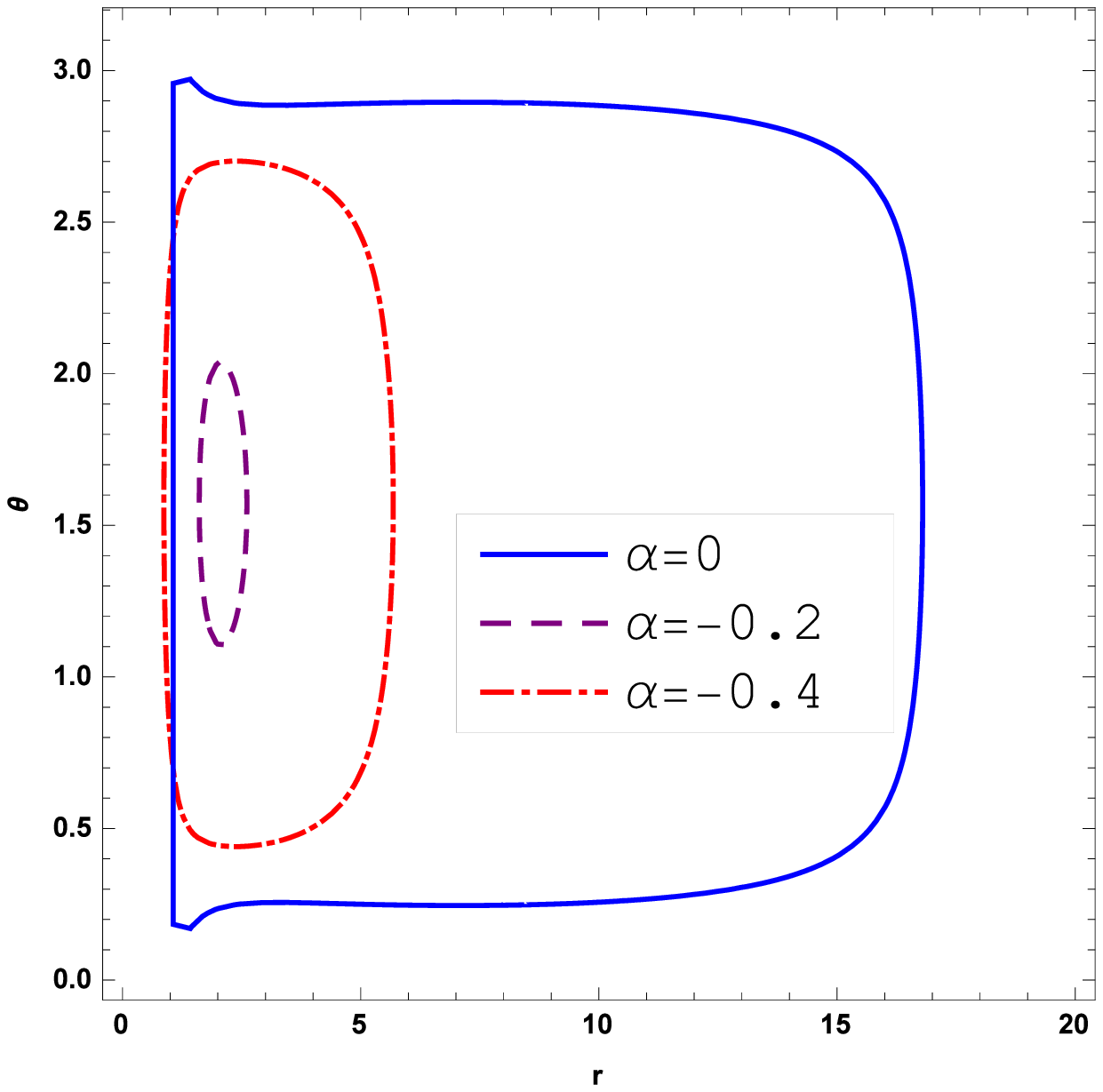}
\caption{Effect of the deformation parameter $\alpha$ on the particle's motion region with $ E=0.94$ and $L=0.8M$. The left and right panels present the boundary of the particle's motion of the particle with different $\alpha$ for fixed $a=0.998$. }\label{fig0}
\end{figure}
\begin{figure}[ht]
\includegraphics[width=5.2cm ]{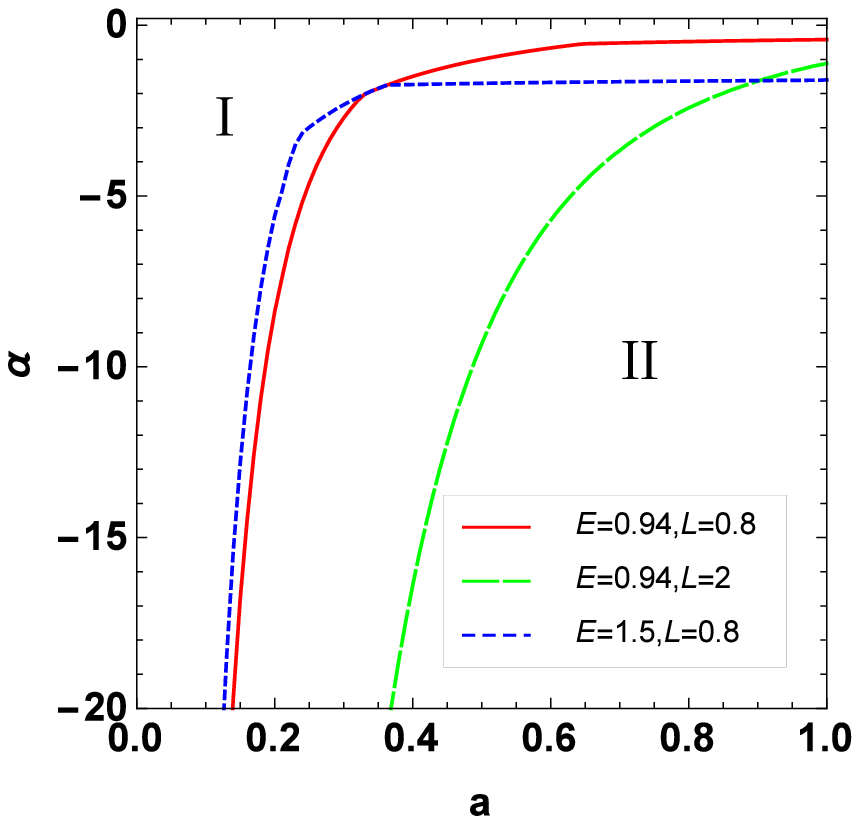}\;\;\;\includegraphics[width=5.2cm ]{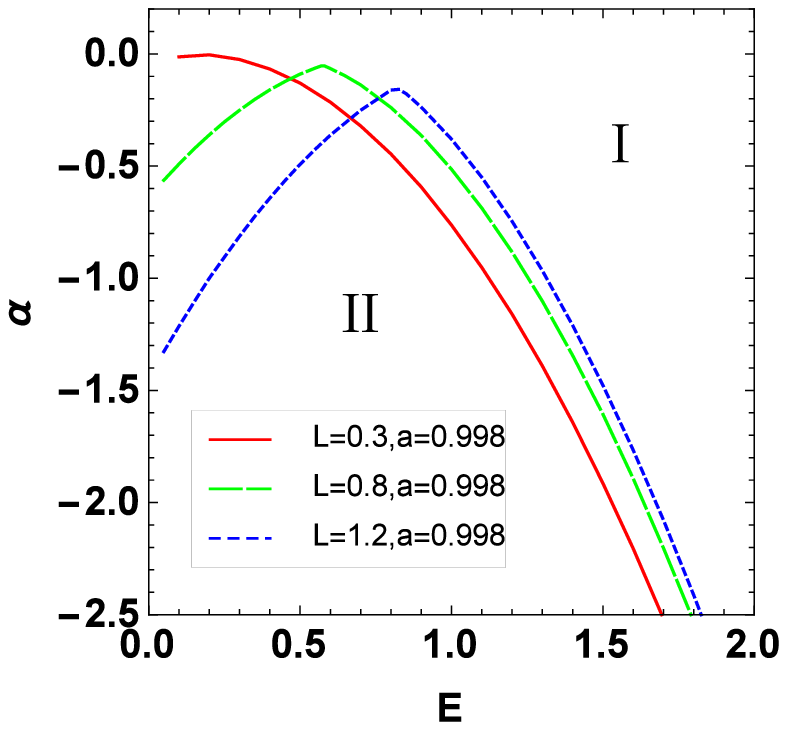}\;\;\;\includegraphics[width=5.0cm ]{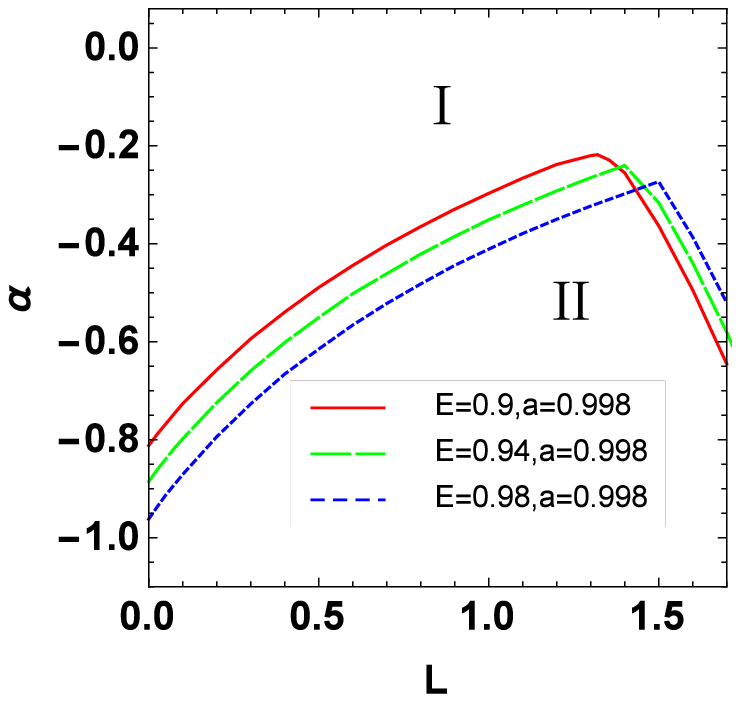}
\caption{Dependence of the threshold value $\alpha_c$ on the parameters $a$, $E$, and $L$ in the disformal rotating Kerr black-hole spacetime. Regions I and II denote the regions where the motion of the given particle exists and does not exist, respectively, in the corresponding parameter panels.} \label{figsa0}
\end{figure}

In principle, the choice for these parameters and initial conditions of the particle is arbitrary. For convenience, this study set the particle's parameters $ E=0.94$ and $L= 0.8M$ and probed the chaotic motion of the chosen particle in the disformal Kerr black-hole spacetime (\ref{metric}).
First, the range of black-hole parameters where the particle motion is allowed must be analyzed. Eqs. (\ref{motion3})-(\ref{motion4}) show $\tilde{g}_{r r} \geq 0$ and $g_{\theta \theta} \geq 0$. Thus, the boundary of the particle's motion region is determined by the condition that the effective potential $V_{eff}\left(r, \theta ; E, L\right)=0$. Fig.\ref{fig0} shows the boundary of the particle's motion region with $E=0.94$ and $L=0.8M$ in the disformal Kerr black-hole spacetime (\ref{metric}) with the spin parameter $a=0.998$. The allowed motion region of the given particle decreases with the decrease of the deformation parameters $\alpha$. Especially, as $\alpha$ decreases down to a threshold value $\alpha_c=-0.42195$, the allowed particle's motion region vanishes, which means that a particle's motion of given $E$ and $L$ exists only for $\alpha>\alpha_c$. Moreover, Fig. \ref{figsa0} numerically presents the dependence of $\alpha_c$ on parameters $a$, $E$, and $L$ in the disformal rotating Kerr black-hole spacetime. For the fixed $E$ and $L$, $\alpha_c$ increases with the spin parameter. As the spin parameter $a$ becomes zero, the $\alpha_c$ approaches negative infinity. For fixed $a=0.998$, $\alpha_c$ monotonously decreases with the particle's energy $E$ for the smaller $L$, but it first increases and then decreases with $E$ for the larger $L$. For the fixed $E$, with increasing angular momentum $L$ of the particle, $\alpha_c$ first increases and then decreases.

\begin{figure}[ht]
\includegraphics[width=14cm ]{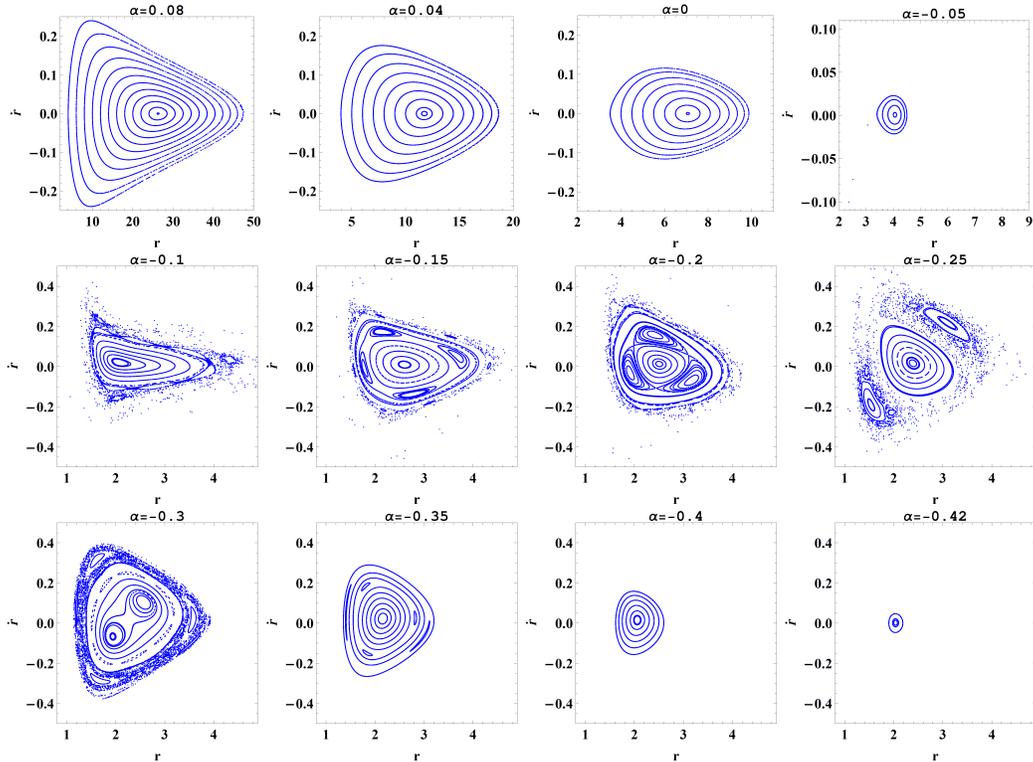}\;\;\;\;
\caption{Change of the Poincar\'e section ($\theta=\frac{\pi}{2}$) with the deformation parameter $\alpha$ for the motion of timelike particles in disformal Kerr black-hole spacetime with the fixed parameters $ a=0.998$, $ E=0.94$, and $L= 0.8M$.}\label{fig1}
\end{figure}

It is well known that the Poincar\'e section is an effective tool to identify chaotic motion, which projects trajectories of a continuous dynamic system on a given hypersurface with a pair of conjugate variables in the phase space. The intersection point distribution in the Poincar\'e section \cite{pjl} reflects the intrinsic dynamic properties of particles' motions. For example, the periodic and quasi-periodic motions correspond to a finite number of points and a series of close curves in the Poincar\'e section, respectively. However, chaotic motions correspond to strange patterns of dispersed points with complex boundaries. Fig. \ref{fig1} shows the change in the Poincar\'e section ($\theta=\frac{\pi}{2}$) with the deformation parameter $\alpha$ for the timelike particle's motion in the disformal Kerr black-hole spacetime (\ref{metric}) with fixed parameters $ a=0.998$, $ E=0.94$, and $L= 0.8M$. The characteristics of the particle's motion depend on the sign of the deformation parameter $\alpha$. For positive $\alpha$, the motion is regular and orderly, and the particle's motion region increases with $\alpha$. For negative $\alpha$, with the increase of the absolute value of $\alpha$, the particle's motion region first decreases, and the chaotic motion appears gradually, and the corresponding particle's motion region increases. With the further increase in $|\alpha|$, the chaotic motion of particles vanishes, and the particle's motion region decreases. Finally, as $\alpha$ decreases to a threshold value $\alpha_c=-0.42195$, the particle's motion region disappears as the motion of particles with given $E=0.94M$ and $L=0.8M$ does not exist as shown in the previous analysis.
Moreover, this study also shows the effect of the spin parameter $a$ on the particle motion for the fixed parameter $\alpha=-0.28$.

\begin{figure}[ht]
\includegraphics[width=15cm ]{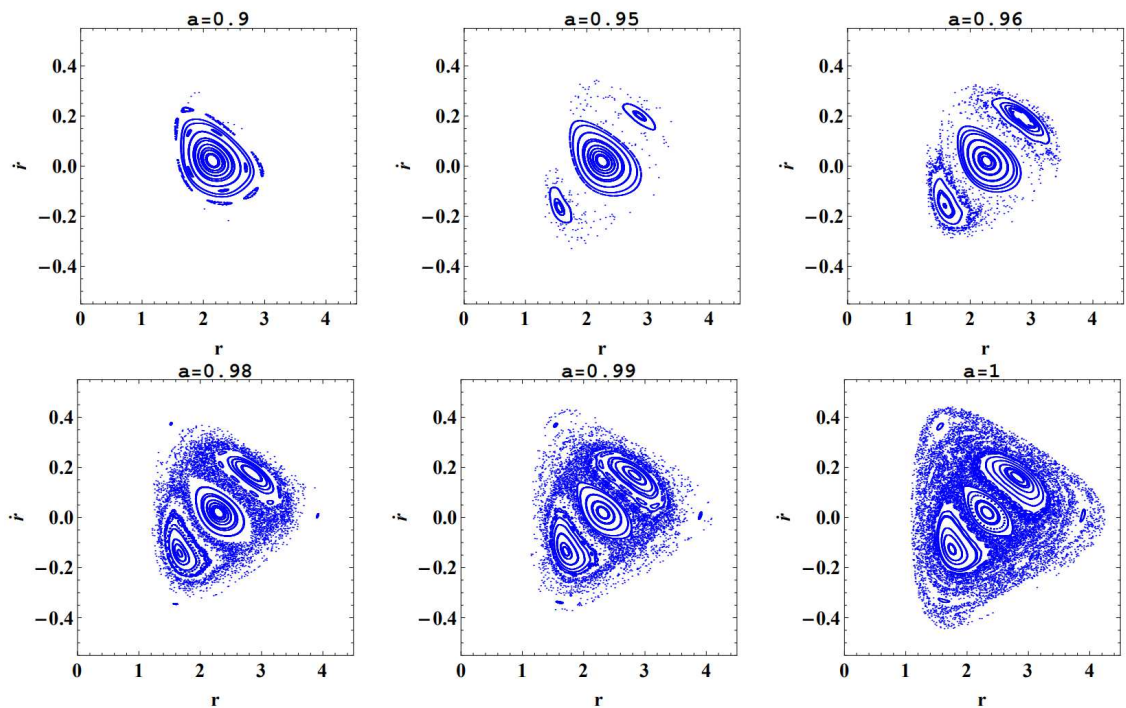}\;\;\;\;
\caption{Change in the Poincar\'{e} section ($\theta=\frac{\pi}{2}$) with the spin parameter $a$ for the motion of timelike particles in disformal Kerr black-hole spacetime with the fixed parameters $\alpha=-0.28$, $E=0.94$, and $L=0.8M$.}\label{fig2}
\end{figure}

Fig. \ref{fig2} shows that the spin parameter $a$ increases the chaotic motion region for the chosen particle.
The spin parameter $a\geq 0.9$ is selected. The main reason is as $\alpha=-0.28$, the motion of particles with given $E=0.94M$ and $L=0.8M$ exists only in the case of the rapidly rotating black hole.
The chaos in the timelike particle's motion in the disformal Kerr black-hole spacetime (\ref{metric}) is detected by analyzing the Lyapunov exponent of the dynamic system. The Lyapunov exponent \cite{Tancredi,Wu5,Wu51,Wu52} describes the growth or decline rate of the deviation vector $\Delta X$ between two nearby trajectories, reflecting actually whether the motion is highly sensitive to the initial conditions. In general relativity, the largest Lyapunov exponent is defined as
\begin{equation}
\centering
L E=\lim _{\tau \rightarrow+\infty}\chi(\tau)=\lim _{\tau \rightarrow+\infty} \frac{1}{\tau} \ln \frac{|\Delta X\left(\tau\right)|}{|\Delta X_{0}|},
\end{equation}
where the length $|\Delta X(\tau)|\equiv\sqrt{\left|g_{\mu \nu} \Delta x^{\mu} \Delta x^{\nu}\right|}$ and
$\Delta x^{\alpha}(\tau)$ is the deviation vector between two nearby trajectories at proper time $\tau$.
\begin{figure}[ht]
\includegraphics[width=8cm ]{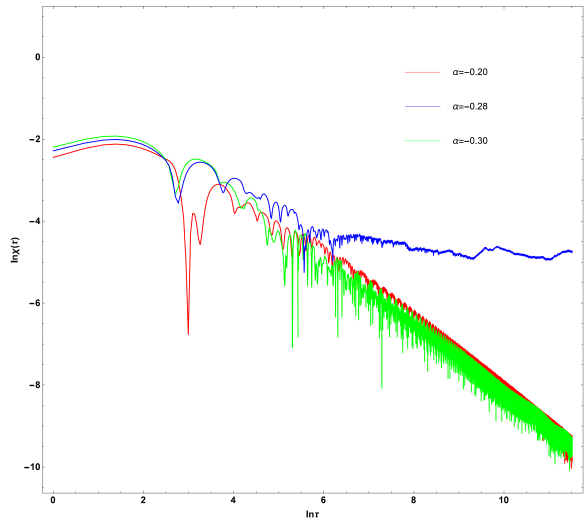}\;\;\;\;
\caption{ Lyapunov exponents with different $\alpha$ for the particle with the parameters $E=0.94$ and $L=0.8$ and the initial conditions $\{$ $r(0)=3.5$; $\dot{r}(0)=0$; and $\theta(0)=\frac{\pi}{2}$$\}$. Here, $a=0.998$. }\label{fig3}
\end{figure}
In practical numerical calculations, it is impossible to take $\tau \rightarrow+\infty$. However, the curve $\ln\chi(\tau)$ versus $\ln\tau$ can be plotted in a longer range in a proper time. If the curve has a constant negative slope, the system is regular. If it presents an inflection of the slope that comes close to $0$ and the plot converges to a certain value of $\chi(\tau)$, then the system is chaotic \cite{Tancredi,Wu5,Wu51,Wu52}.
Fig. \ref{fig3} presents the change $\ln\chi(\tau)$ of $\ln\tau$ with different $\alpha$ for the particle with parameters $E=0.94$, $L=0.8$, and the initial conditions $\{$ $r(0)=3.5$; $\dot{r}(0)=0$; and $\theta(0)=\frac{\pi}{2}$$\}$ in the disformal Kerr black-hole spacetime (\ref{metric}) with $a=0.998$. As $\alpha=-0.2$ or $-0.3$, the curve has a negative constant slope, which means that the motion of the particle is orderly. As $\alpha=-0.28$, it presents an inflection of the slope that comes close to $0$ and converges to a certain value of $\chi(\tau)$, implying that the largest Lyapunov exponent is positive and the corresponding motion is chaotic. Fig. \ref{fig4} also presents the Poincar\'{e} section for the particle motion in the above three cases.

\begin{figure}[ht]
\includegraphics[width=5cm ]{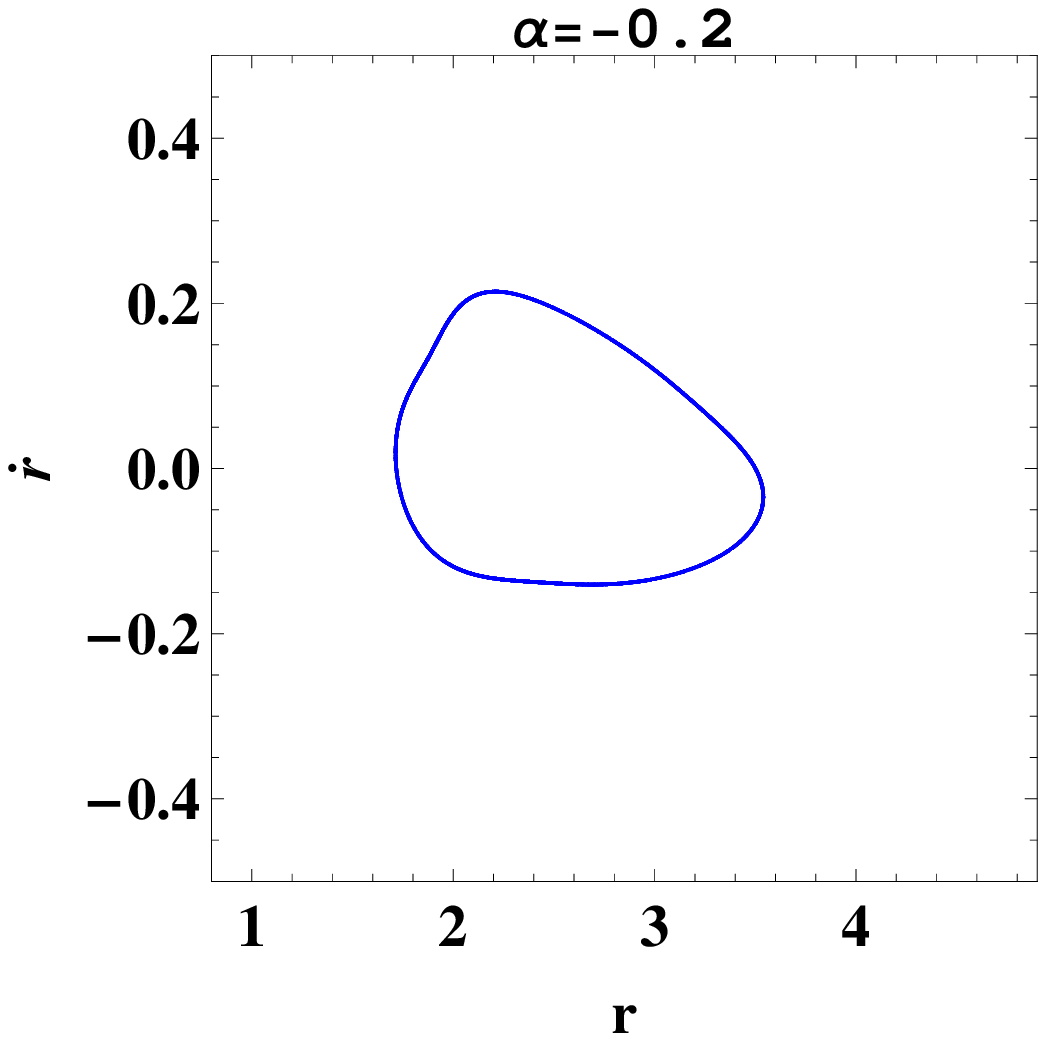}\;\;\;\;
\includegraphics[width=5cm ]{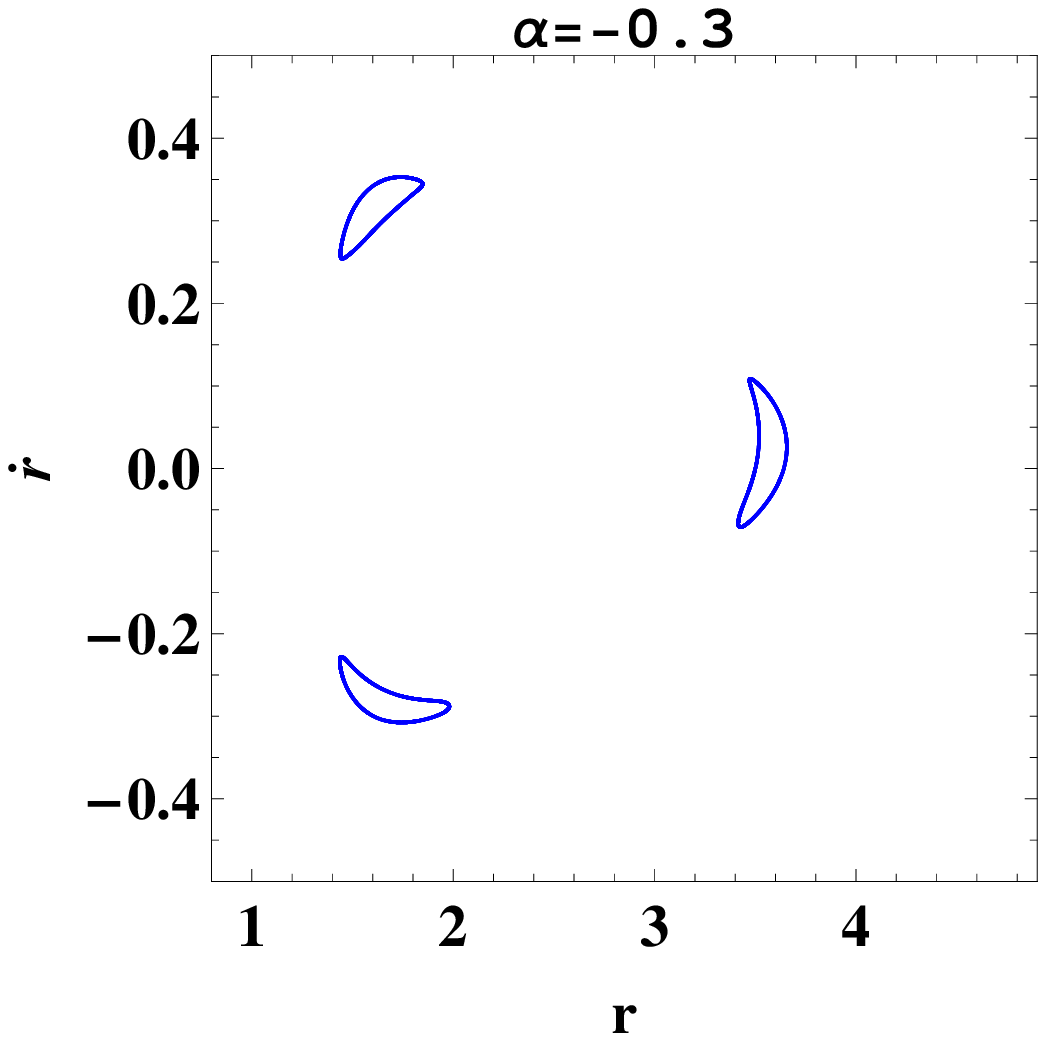}\;\;\;\;
\includegraphics[width=5cm ]{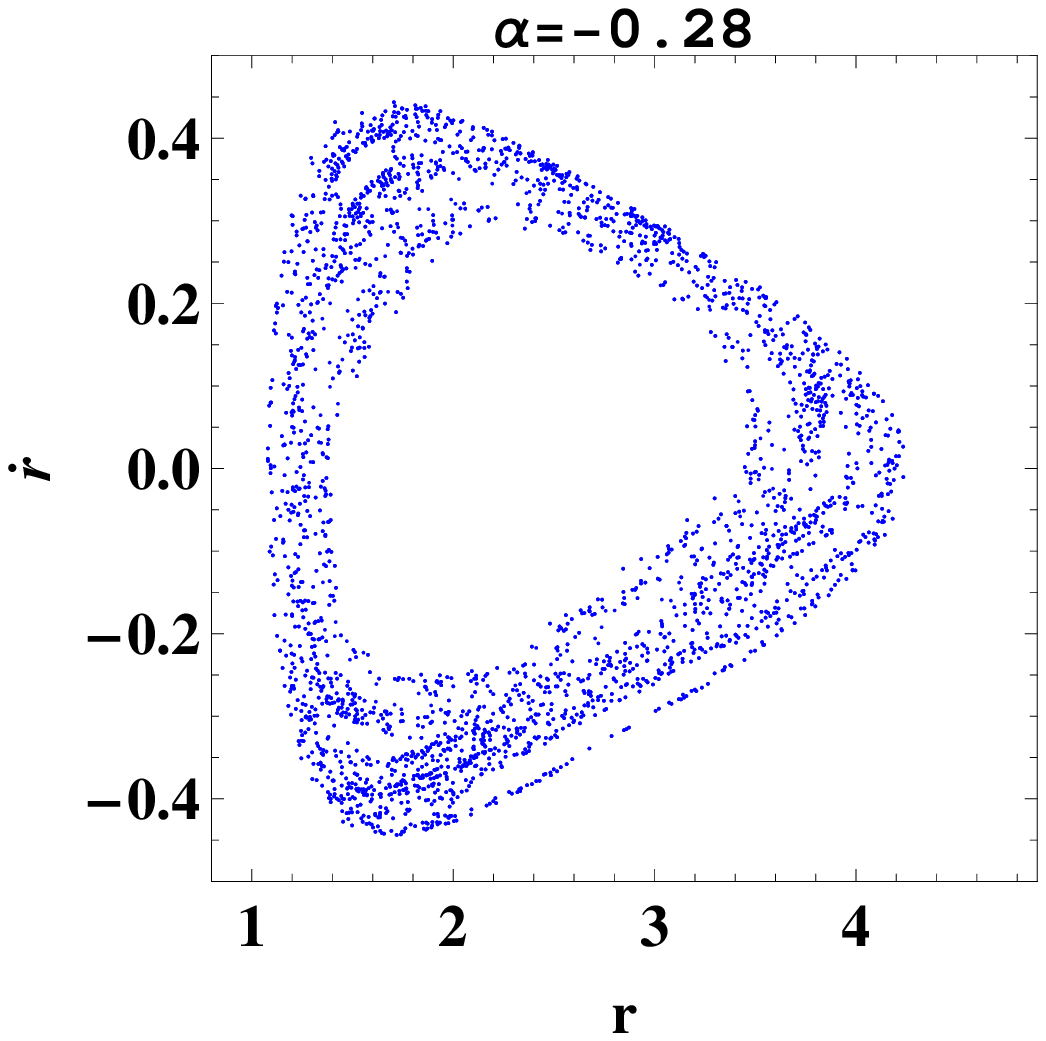}\;\;\;\;
\caption{ Poincar\'e surface of section $(\theta=\frac{\pi}{2})$ with different $\alpha$ for the particle motion in Fig.\ref{fig3}. }\label{fig4}
\end{figure}

In $\alpha=-0.2$, the phase path of the particle motion is a quasi-periodic Kolmogorov--Arnold--Moser (KAM) torus, and the corresponding motion is regular. As $\alpha=-0.3$, there is a chain of islands composed of three secondary KAM tori belonging to the same trajectory. However, as $\alpha=-0.28$, the KAM torus is broken, and there are many discrete points distributed randomly in the Poincar\'{e} section, and the corresponding motion is chaotic. The dynamic properties of the particle with the chosen initial conditions agree with those obtained by analyzing the Lyapunov exponent in Fig.\ref{fig3}.

\begin{figure}[ht]
\includegraphics[width=16cm ]{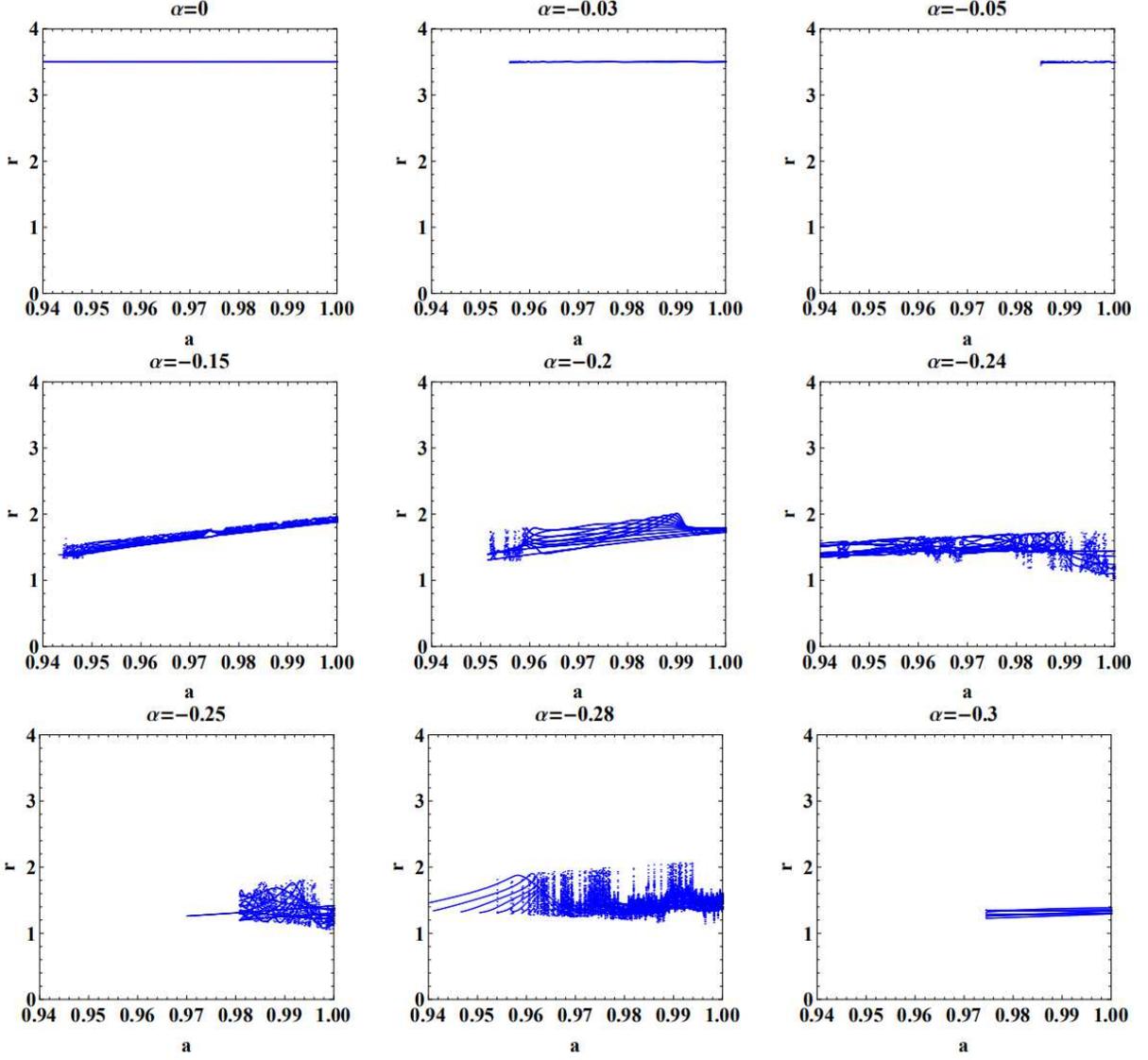}\;\;\;\;
\caption{ Bifurcation with the deformation parameter $\alpha$ for the motion of the timelike particle with the parameters $E=0.94$, $L=0.8$, and the initial conditions $\{$ $r(0)=3.5$; $\dot{r}(0)=0$; and $\theta(0)=\frac{\pi}{2}$$\}$ in the disformal Kerr black-hole spacetime (\ref{metric}).}\label{fig6}
\end{figure}

\begin{figure}[ht]
\includegraphics[width=16cm ]{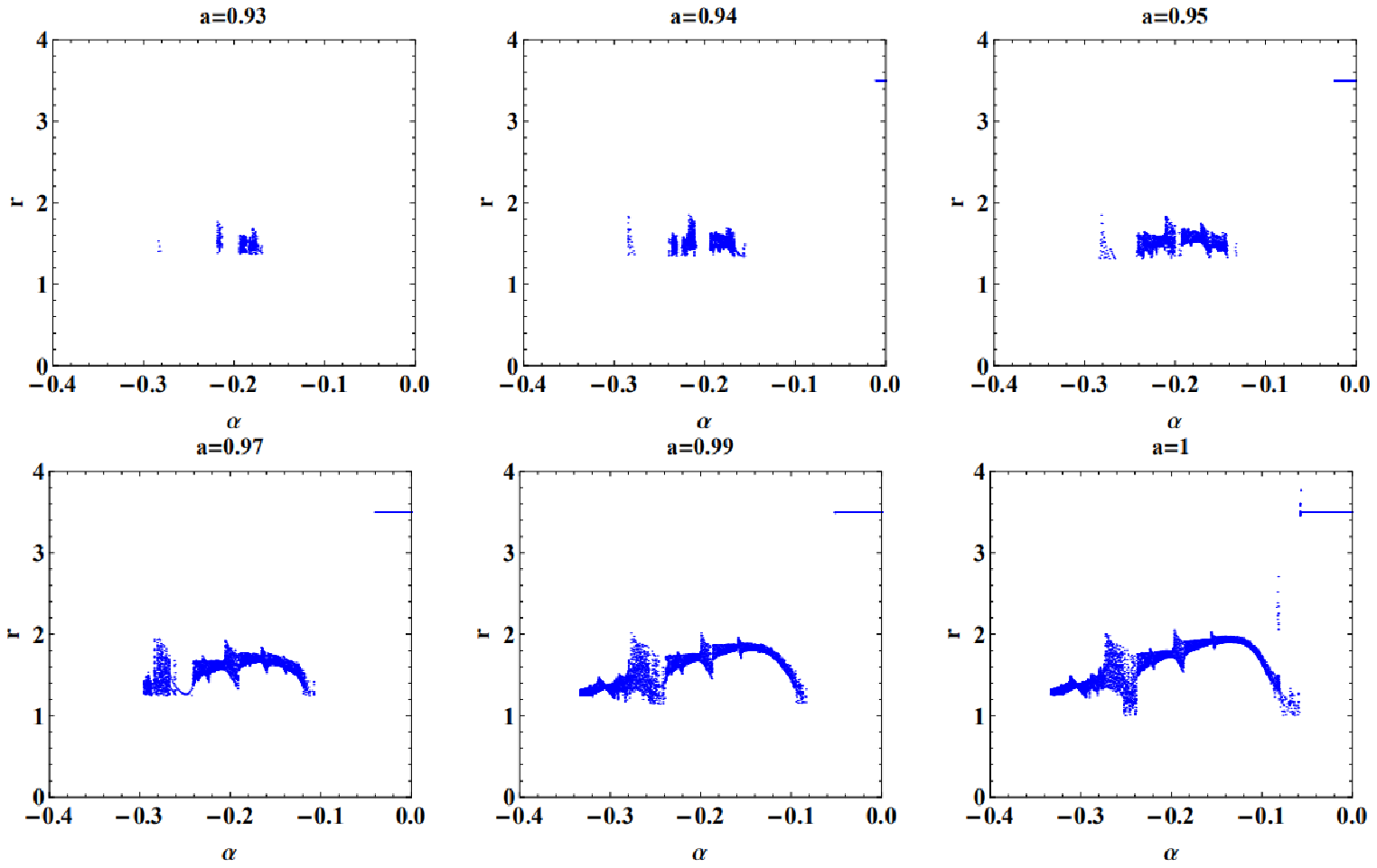}\;\;\;\;
\caption{ Bifurcation with the spin parameter $a$ for the motion of the timelike particle with the parameters $E=0.94$, $L=0.8$, and the initial conditions $\{$ $r(0)=3.5$; $\dot{r}(0)=0$; and $\theta(0)=\frac{\pi}{2}$$\}$ in the disformal Kerr black-hole spacetime (\ref{metric}).}\label{fig7}
\end{figure}

The dependence of the dynamic behaviors of the system on the black-hole parameters can also be visualized in the form of a bifurcation diagram. Figs. \ref{fig6} and \ref{fig7} plot the bifurcation diagram of the radial coordinate $r(\tau)$ with the deformation parameter $\alpha$ and the spin parameter $a$ for the particle motion with parameter $E=0.94$, $L=0.8$, and the initial conditions $\{$ $r(0)=3.5$; $\dot{r}(0)=0$; and $\theta(0)=\frac{\pi}{2}$$\}$ in the disformal Kerr black-hole spacetime (\ref{metric}). This study shows the results only for $a> 0.90$, as no stable orbit for the above particle with the chosen initial conditions in the slow rotation black-hole case can be found.
In $\alpha=0$, the radial coordinate $r(\tau)$ is a periodic function, and there is no bifurcation for the dynamic system as the metric (\ref{metric}) reduces to the Kerr one and the corresponding dynamic system of the particle is integrable because the condition (\ref{con4}) is variable-separable in this case.

In $\alpha\neq0$, there obviously exist periodic, chaotic, and escaped solutions that depend on the deformation parameter $\alpha$ and the spin parameter $a$. Moreover, as the parameters $\alpha$ and $a$ change, the motion of particles transforms among single-periodic, multiple-periodic, and chaotic motions. It means that the effects of the parameters $\alpha$ and $a$ on the particle's motion are very complex, which is the typical feature of the bifurcation diagram in the usual chaotic dynamic system. Moreover, in Figs. \ref{fig6} and \ref{fig7}, the range of $\alpha$ for the existence of the oscillation solution increases with the spin parameter $a$. With the decrease in $\alpha$, the range of $a$ for the existence of the oscillation solution becomes more complicated. This means that the dynamic behavior of timelike particles in the disformal Kerr black-hole spacetime (\ref{metric}) becomes richer than that in the usual Kerr black-hole case.

\section{Summary}
This study investigated the dynamic behaviors of the motion of timelike particles in the disformal rotating black-hole spacetime with noncircularity in quadratic DHOST theories. First, the motion of particles only exists, as the deformation parameter $\alpha$ is larger than a certain threshold value $\alpha_c$. For the particle with given $E$ and $L$, threshold value $\alpha_c$ increases with the spin parameter $a$. As the spin parameter $a$ tends to be zero, the $\alpha_c$ approaches negative infinity. Moreover, the particle's motion characteristics depend on the sign of the deformation parameter $\alpha$. For positive $\alpha$, the motion is regular and orderly, and the particle's motion region increases with $\alpha$. For negative $\alpha$, with the increase in the absolute value of $\alpha$, the particle's motion region first decreases, the chaotic motion appears gradually, and the corresponding particle's motion region increases. With the further increase in $|\alpha|$, the chaotic motion of particles vanishes, and the particle's motion region decreases. Finally, as $\alpha$ decreases to a threshold value $\alpha_c$, the particle's motion region disappears, and the particle's motion with given $E$ and $L$ does not exist in the black-hole spacetime with given $a$. The presence of chaos in the particle's motion means that the dynamic behavior of timelike particles in the disformal rotating black-hole spacetime with noncircularity becomes richer than that in the usual Kerr black-hole case.

\section{\bf Acknowledgments}

This work was  supported by the National Natural Science
Foundation of China under Grant No.11875026, 11875025, 12035005, and 2020YFC2201403.

\end{document}